\documentclass[conference]{IEEEtran}
\usepackage{caption}
\usepackage{subfigure}
\usepackage{graphicx}
 \usepackage{float} 
\IEEEoverridecommandlockouts
\usepackage{cite}
\usepackage{amsmath,amssymb,amsfonts}
\usepackage{algorithmic}
\usepackage{graphicx}
\usepackage{textcomp}
\usepackage{xcolor}
\def\BibTeX{{\rm B\kern-.05em{\sc i\kern-.025em b}\kern-.08em
    T\kern-.1667em\lower.7ex\hbox{E}\kern-.125emX}}
\begin{document}

\title{COVID-19 Detection in Chest X-ray Images Using Swin-Transformer and Transformer in Transformer  \\
}

\author{\IEEEauthorblockN{1\textsuperscript{st} Juntao Jiang}
\IEEEauthorblockA{\textit{College of Control Sceince and Engineering} \\
\textit{Zhejiang University}\\
Hangzhou, China \\
juntaojiang@zju.edu.cn}
\and
\IEEEauthorblockN{2\textsuperscript{nd} Shuyi Lin}
\IEEEauthorblockA{\textit{Khoury College of Computer Sciences} \\
\textit{Northeastern University}\\
Boston, United States \\
lin.shuyi@northeastern.edu}

}

\maketitle

\begin{abstract}
The Coronavirus Disease 2019 (COVID-19) has spread globally and caused serious damage. Chest X-ray images are widely used for COVID-19 diagnosis, and the Artificial Intelligence method can increase efficiency and accuracy. In the Challenge of Chest XR COVID-19 detection in Ethics and Explainability for Responsible Data Science (EE-RDS) conference 2021, we proposed a method that combined Swin Transformer and Transformer in Transformer to classify chest X-ray images as three classes: COVID-19, Pneumonia, and Normal (healthy)  and achieved 0.9475 accuracies on the test set.
\end{abstract}


\section{Introduction}
The Coronavirus Disease 2019 (COVID-19), caused by the severe acute respiratory syndrome coronavirus-2 (SARS-CoV-2, 2019-nCoV), has become a global pandemic and brought unprecedented damage seriously worldwide. As Chest X-ray tests typically have a high sensitivity diagnosis of COVID-19\cite{b1,b2}, Chest X-ray images can be used for not only following up on the effects of COVID-19 on lung tissue but also for early detection of COVID-19; thus the immediate isolation and treatment for the suspected can be achieved.\par
Past years witnessed the growing use of AI techniques, especially deep-learning-based methods, in disease detection on chest X-ray images \cite{b3,b4,b5,b6}, successfully increasing the accuracy and efficiency for early diagnosis. After the pandemic outbreak, many approaches have also been proposed to detect COVID-19 in chest X-ray images. Shervin Minaee et al.\cite{b8} trained four state-of-the-art convolutional networks for COVID-19 detection on a dataset of around 5000 X-ray images and achieved higher than 90 percent of sensitivity and specificity rate. Asif IqbalKhan et al.\cite{b9} designed a deep convolutional neural network model based on Xception architecture to classify Normal, Pneumonia-bacterial, Pneumonia-viral, and COVID-19 chest X-ray images. Linda Wang et al.\cite{b10 } introduced a deep convolutional neural network called COVID-Net to detect COVID-19 cases from chest X-ray images, which is open source and available to the general public. Rachna Jain et al. \cite{b11} used Inception net V3, XCeption net and ResNeXt to classify. Sanhita Basu et al. \cite{b12} gave a new concept called domain extension transfer learning (DETL) with the pre-trained deep convolutional neural network on a related large chest X-Ray dataset. We can conclude that the fundamental idea for most current methods is based on multi-convolutional neural networks and transfer learning from a large dataset. \par
Special attention should be paid to the fact that Transformer methods have recently outperformed convolutional neural networks in some datasets of different scenes. Swin Transformer\cite{b13} and Transformer in Transformer\cite{b14 } are successful works in adapting Transformer from language to vision and achieved state-of-the-art in different tasks. Still, we haven't seen a lot of applications in chest X-ray image classification and COVID-19 detection. This paper is a technical report in Chest XR COVID-19 Detection Challenge\cite{b140 }, a part of Ethics and Explainability for Responsible Data Science (EE-RDS) conference. We combine Swin-transformer and Transformer in Transformer to classify chest X-ray images into three classes, COVID-19, Pneumonia, and Normal (healthy), on the dataset offered by the challenge organizers and achieve 0.9475 accuracies on the test set.

\section{Introduction to the Dataset and the Task}
\begin{figure}
\includegraphics[scale=0.3]{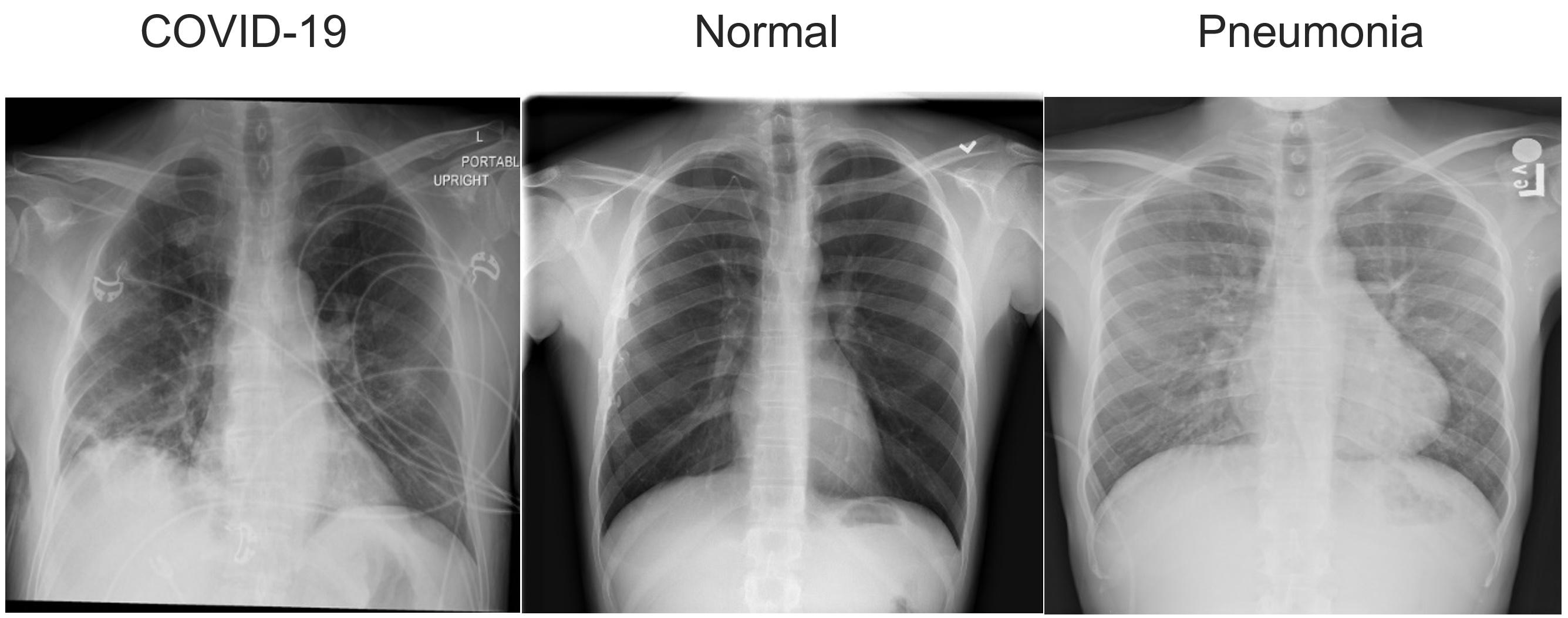}
\caption{Example Chest X-ray Images for classes of COVID-19, Normal and Pneumonia} \label{fig1}
\end{figure}
\begin{figure*}[htbp]
\includegraphics[scale=0.6]{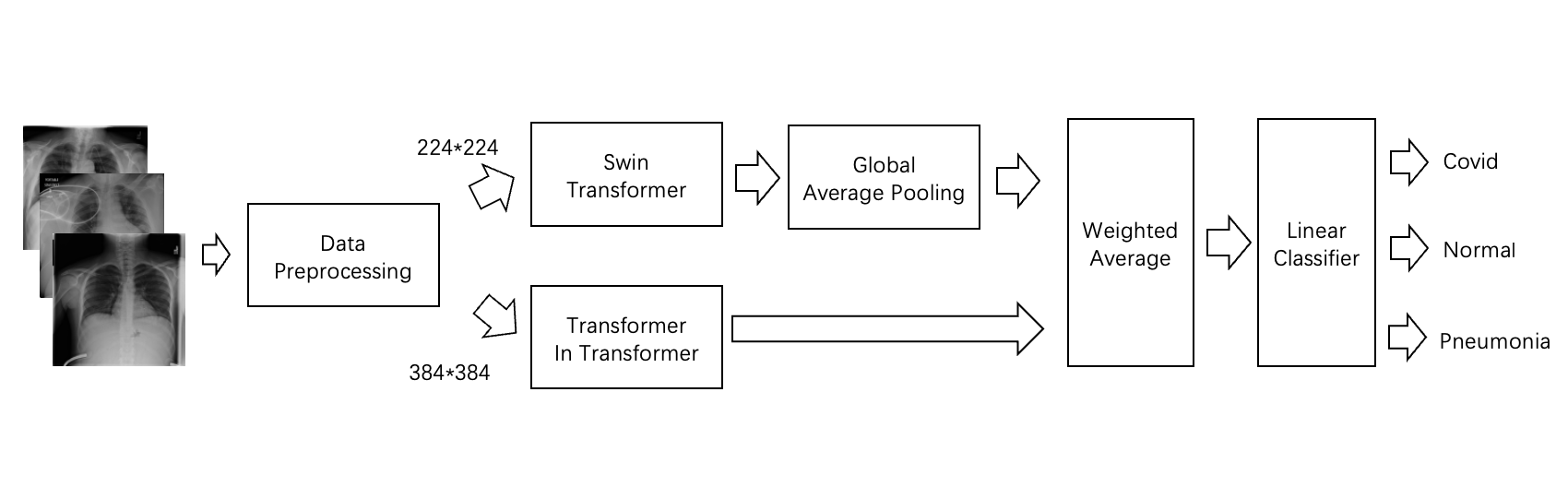}
\caption{Workflow for Chest X-ray Images Classification with Swin Transformer and Transformer in Transformer} \label{fig1}
\end{figure*}

\begin{figure*}[htbp]
\includegraphics[scale=0.6]{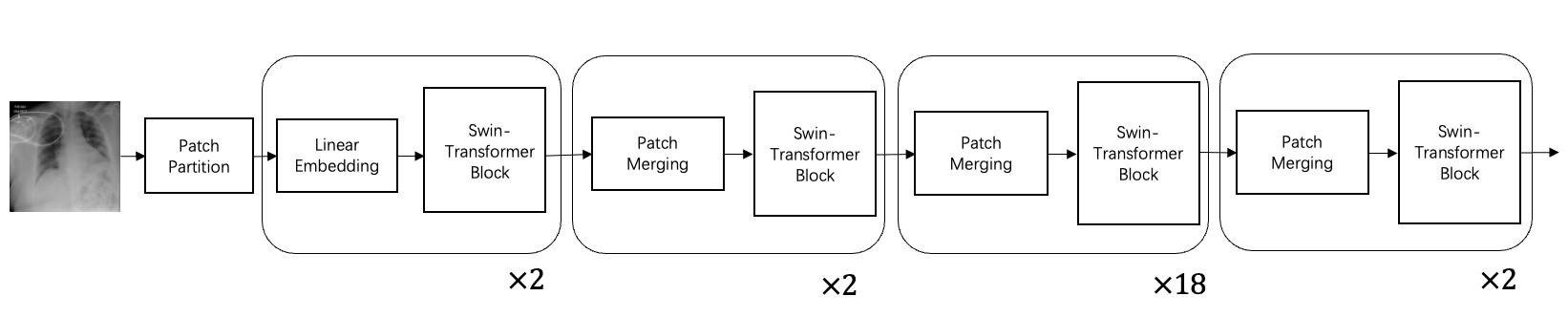}
\caption{The architecture of Swin-Transformer (Swin-B)} \label{fig1}
\end{figure*}
\begin{figure}[htbp]
\includegraphics[scale=0.4]{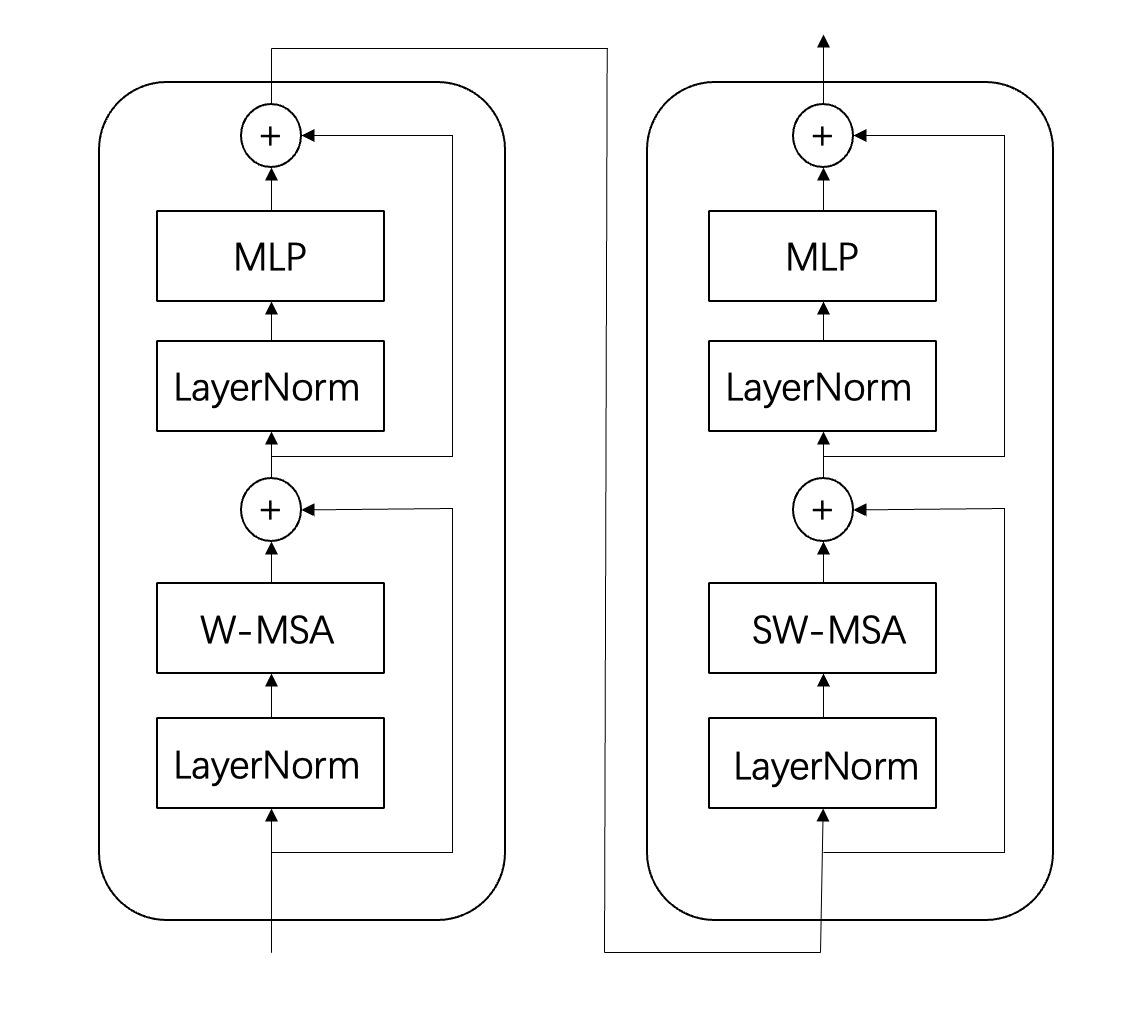}
\caption{Two successive Swin Transformer Blocks, where W-MSA and SW-MSA are multi-head self-attention modules with regular and shifted windowing configurations, respectively.} \label{fig1}
\end{figure}
\begin{figure*}[htbp]
\includegraphics[scale=0.8]{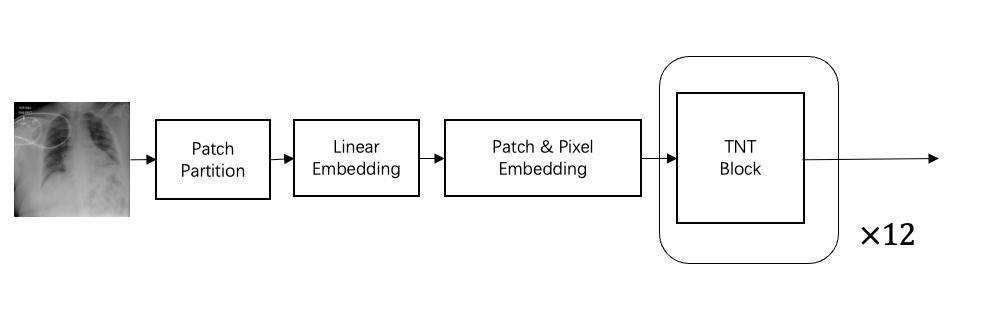}
\caption{The architecture of Transformer in Transformer (TNT-S)} \label{fig1}
\end{figure*}
\begin{figure}[htbp]
\includegraphics[scale=0.6]{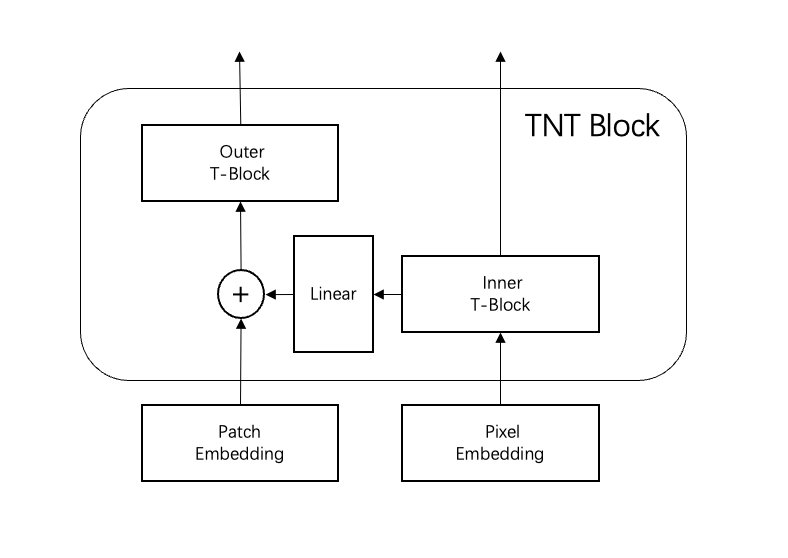}
\caption{The architecture of TNT block} \label{fig1}
\end{figure}
The dataset of this challenge contains three parts, which are used to train, validate and test, respectively:
\begin{itemize}
\item Train = 17,955 chest X-ray images
\item validation = 3,430 chest X-ray images
\item test = 1,200 chest X-ray images
\end{itemize}
This dataset has three types of chest X-ray images: COVID-19, Pneumonia, and Normal (healthy). The example of three classes are shown in Figure 1.\par

Our task is to design an algorithm to auto-classify chest X-ray images into these three classes. \par

\section{Methods}
\begin{figure*}[htbp]
\includegraphics[scale=0.6]{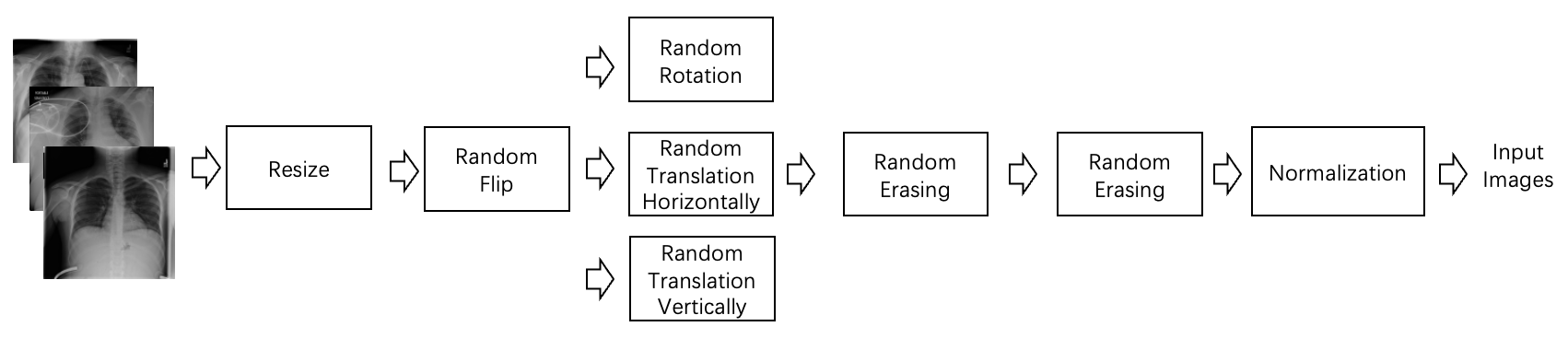}
\caption{Data Preprocessing in Training} \label{fig1}
\end{figure*}
\begin{figure}[htbp]
\includegraphics[scale=0.6]{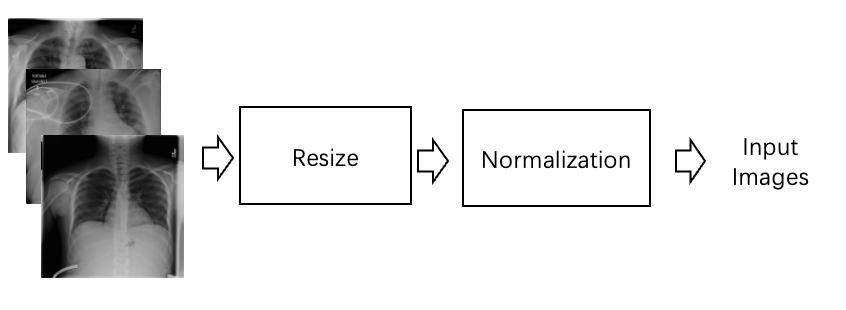}
\caption{Data Preprocessing in Testing} \label{fig1}
\end{figure}
The workflow of our classification method is shown in Figure 3. We trained the Swin Transformer and Transformer in Transformer separately and did results ensemble using weighted average methods. 

\subsection{Data Preprossing} 
The workflow of our Preprossing method for training is shown in figure 3, and the workflow of our Preprossing method for training is shown in figure 4.\par

Firstly we resized the images to a certain scale (224*224 or 384*384), then applied random flipping, then selected a policy from rotation, translation horizontally or vertically, and then applied random erasing. Finally, we did the normalization. All these changes are applied with a small probability.\par
Unlike natural images, the overall characteristics of medical images are significant to classify, so we avoid using crop or center crop as data augmentation methods. We only use horizontal flipping, and the rotation or translation augmentation methods are also controlled within a small range. \par
We did not use prepossessing methods related to brightness or contrast to avoid destroying the difference between the three classes.

\subsection{Models} 
\subsubsection{Swin Transformer}Swin Transformer\cite{b13} constructs hierarchical feature maps to better model objects of different sizes and has linear computational complexity to image size. Swin Transformer has different types according to the models' size and we selected Swin-B for the task. The architecture of Swin-B is shown in Figure 3. 

As shown in Figure 4, a Swin Transformer block utilizes a shifted window-based Multi-head Self-Attention (SW-MSA) module, followed by a 2-layer MLP with GELU nonlinearity in between. A LayerNorm layer is applied
before each MSA module and MLP, and a residual connection is applied.

\subsubsection{Transformer in Transformer}
Transformer in Transformer (TNT)\cite{b14}, which regards the local patches as "visual sentences" and then presents to further divide them into smaller patches as "visual words," and features of both words and sentences will be aggregated to improve the representation ability, thus the patch-level and pixel-level representations can be modeled.\par
Transformer in Transformer has different types according to the models' size, and we selected the Transformer in Transformer small for the task. The larger models should have better performance, but we failed to try them due to the time limitation.  The architecture of Transformer in Transformer small (TNT-S) is shown in Figure 5, which consists of 12 TNT blocks. The architecture of a TNT block is shown in Figure 6.

\subsection{Training Setting}
Training settings. Our implementation is based on PyTorch \cite{b15}, mmclassification\cite{b150}, and one NVIDIA Tesla-V100 GPU is used for training. Model training is done using the AdamW, and the initial learning rate, batch size, and weight decay are set to 0.001, 64, and 0.05, respectively. For the learning rate scheduler, cosine annealing \cite{b16} and warmup \cite{b17} are used. We also applied label smoothing in the loss.  \par
In training for Swin Transformer, the input image size is 224*224. For Transformer in Transformer, the input image size is 384*384. The design is due to the limitation of computation power, and also for convenience to use the pre-trained model directly from ImageNet \cite{b18}.
\subsection{Ensemble}
We obtain the probabilities for each class from two models, then calculate the weighted average for these probabilities. Finally, a linear classifier was used to get the final results for classification.

\section{Experiments and Results}
We train the two models on the training set and validate them on the validation set. Then we select the best model on the validation set to infer on the test set. \par
Due to the time limitation, the experiments are not enough for thorough comparison but may still give some information for further research. The experimental results are shown in Table 1.

More specifically, our final results are:
\begin{itemize}
\item Accuracy Score: 0.9475,
\item Sensitivity Score: 0.9475
\item Specificity Score: 0.9509
\end{itemize}

Our results rank ten on the leaderboard for this challenge.
\begin{table}
\caption{Results for Image Classification }\label{tab1}
\begin{tabular}{|l|l|l|}
\hline
Models  &  Weights & Accuracy\\
\hline
Swin Transformer &  & 0.9467\\
Swin Transformer, TNT & 1:1 & 0.9467\\
Swin Transformer, TNT & 2:1&
0.9475\\
\hline
\end{tabular}
\end{table}
\section{Conclusion}
In this paper, we applied Swin Transformer and Transformer in Transformer to classify the chest X-ray images in In the Challenge of Chest XR COVID-19 detection. We did a model ensemble by using the weighted average method. We  achieved  0.9475 accuracies on the test set and ranked ten on the leaderboard.

\vspace{12pt}


\begin{thebibliography}{00}
\bibitem{b1} Heshui Shi, Xiaoyu Han, Nanchuan Jiang, Yukun Cao, Osamah Alwalid, Jin Gu, Yanqing Fan, Chuansheng Zheng,Radiological findings from 81 patients with COVID-19 pneumonia in Wuhan, China: a descriptive study, The Lancet Infectious Diseases, 2020, Pages 425-434
\bibitem{b2} T. Ai, Z. Yang, H. Hou, C. Zhan, C. Chen, W. Lv, Q. Tao, Z. Sun, L. Xia
Correlation of chest CT and RT-PCR testing in coronavirus disease 2019 (COVID-19) in China: a report of 1014 cases, Radiology (2020), 10.1148/radiol.2020200642
\bibitem{b3} Y. Dong, Y. Pan, J. Zhang, W. Xu, Learning to read chest X-ray images from 16000+ examples using CNN
Proceedings, 2017 IEEE/ACM International Conference on Connected Health: Applications, Systems and Engineering Technologies, IEEE (2017), pp. 51-57
\bibitem{b4} P. Rajpurkar, J. Irvin, R.L. Ball, K. Zhu, B. Yang, H. Mehta, ..., B.N. Patel, Deep learning for chest radiograph diagnosis: A retrospective comparison of the CheXNeXt algorithm to practicing radiologists, PLoS Medicine, 15 (11) (2018)
\bibitem{b5} V. Chouhan, S.K. Singh, A. Khamparia, D. Gupta, P. Tiwari, C. Moreira, ..., V.H.C. de Albuquerque, A novel transfer learning based approach for pneumonia detection in chest X-ray images, Applied Sciences, 10 (2) (2020)
\bibitem{b6}C. Liu, Y. Cao, M. Alcantara, B. Liu, M. Brunette, J. Peinado, W. Curioso, TX-CNN: Detecting tuberculosis in chest X-ray images using convolutional neural network 2017 IEEE international conference on image processing (ICIP), IEEE (2017), pp. 2314-2318.
\bibitem{b8} Shervin Minaee, Rahele Kafieh, Milan Sonka, Shakib Yazdani, Ghazaleh Jamalipour Soufi, Deep-COVID: Predicting COVID-19 from chest X-ray images using deep transfer learning, Medical Image Analysis, Volume 65, 2020
\bibitem{b9}Asif Iqbal Khan, Junaid Latief Shah, Mohammad Mudasir Bhat, CoroNet: A deep neural network for detection and diagnosis of COVID-19 from chest x-ray images,
Computer Methods and Programs in Biomedicine, Volume 196,2020
\bibitem{b10}Wang, L., Lin, Z.Q.,Wong, A. COVID-Net: a tailored deep convolutional neural network design for detection of COVID-19 cases from chest X-ray images. Sci Rep 10, 19549 (2020).
\bibitem{b11}Jain, R., Gupta, M., Taneja, S. et al, Deep learning based detection and analysis of COVID-19 on chest X-ray images. Appl Intell 51, 1690–1700 (2021). 

\bibitem{b12}S. Basu, S. Mitra and N. Saha, Deep Learning for Screening COVID-19 using Chest X-Ray Images,2020 IEEE Symposium Series on Computational Intelligence (SSCI), 2020, pp. 2521-2527.

\bibitem{b13}Ze Liu, Yutong Lin, Yue Cao, Han Hu, Yixuan Wei, Zheng Zhang, Stephen Lin, Baining Guo, Swin Transformer: Hierarchical Vision Transformer using Shifted Windows, arXiv:2103.14030
\bibitem{b14}Kai Han, An Xiao, Enhua Wu, Jianyuan Guo, Chunjing Xu, Yunhe Wang, Transformer in Transformer, arXiv:2103.00112
\bibitem{b140}Akhloufi, Moulay A. and Chetoui, Mohamed, Chest XR COVID-19 detection,https://cxr-covid19.grand-challenge.org/,August,2021
\bibitem{b15}Adam Paszke, Sam Gross, Francisco Massa, Adam Lerer, James Bradbury, Gregory Chanan, Trevor Killeen, Zeming Lin, Natalia Gimelshein, Luca Antiga, Alban Desmaison, Andreas Köpf, Edward Yang, Zach DeVito, Martin Raison, Alykhan Tejani, Sasank Chilamkurthy, Benoit Steiner, Lu Fang, Junjie Bai, Soumith Chintala, PyTorch: An Imperative Style, High-Performance Deep Learning Library, NeurIPS 2019
\bibitem{b150}MMClassification Contributors,OpenMMLab's Image Classification Toolbox and Benchmark, https://github.com/open-mmlab/mmclassification, 2020
\bibitem{b16}I. Loshchilov and F. Hutter. SGDR: stochastic gradient descent with warm restarts. ICLR, 2017. 
\bibitem{b17}Kaiming He, Xiangyu Zhang, Shaoqing Ren, Jian Sun, Deep Residual Learning for Image Recognition, arXiv:1512.03385
\bibitem{b18}J. Deng, W. Dong, R. Socher, L. Li, Kai Li and Li Fei-Fei, ImageNet: A large-scale hierarchical image database, 2009 IEEE Conference on Computer Vision and Pattern Recognition, 2009
\end{thebibliography}
\end{document}